\documentclass[fleqn,usenatbib]{mnras}

\usepackage{newtxtext,newtxmath}

\usepackage[T1]{fontenc}
\usepackage{ae,aecompl}

\newcommand{\beq}{\begin{equation}}
\newcommand{\eeq}{\end{equation}}
\newcommand{\beqa}{\begin{eqnarray}}
\newcommand{\eeqa}{\end{eqnarray}}

\usepackage{graphicx}	
\usepackage{amsmath}	
\usepackage{amssymb}	

\title[Theory of pc-GR]
{Predictions of the pseudo-complex theory of Gravity for 
EHT observations- II. Theory and predictions}
\author[P. O. Hess et al.]{
P. O. Hess,$^{1,4}$\thanks{E-mail: hess@nucleares.unam.mx}
Th. Boller,$^{2,4}$
A. M\"uller$^{3}$,
and H. St\"ocker$^{4,5,6}$
\\
$^{1}$Instituto de Ciencias Nucleares, UNAM, Circuito Exterior,
 C.U., A.P. 70-543, 04510, Mexico D.F., Mexico\\
$^{2}$Max-Planck Institute for Extraterrestial Physics, 
Giessenbachstrasse, D-85748 Garching, Germany\\
$^{3}$ Excellence Cluster Universe, Technical University Munich, Giessenbachstrasse 1, 85741 Garching, Germany\\
$^{4}$Frankfurt Institute for Advanced Studies, Johann Wolfgang Goethe
 Universit\"at, Ruth-Moufang-Str.1, 60438 Frankfurt am Main, Germany \\  
$^{5}$Goethe Universit\"at,
Max-von-Laue-Strasse 1, 60438 Frankfurt am Main, Germany \\
$^{6}$GSI Helmholtzzentrum f\"ur Schwerionenforschung GmbH, Planckstrasse 1, 64291 Darmstadt, Germany
}
\date{Accepted 2019 March 17. Received 2019 March 7; in original form 2018 December 26.}
\pubyear{2019}
\begin{document}
\label{firstpage}
\pagerange{\pageref{firstpage}--\pageref{lastpage}}
\maketitle
\begin{abstract}
We present a resum\'e on the modified theory of gravity, called {\it pseudo-complex General Relativity} (pc-GR).
It is the second in a series of papers, where the first one 
(Boller et al. 2019, referred to as paper I)
discussed the observational consequences of pc-GR.
In this paper, we concentrate on the underlying theory. 
PC-GR involves an algebraic extension of the standard theory of GR 
and it depends on two phenomenological parameters. 
An element included in pc-GR that is not present in standard GR is
the energy-momentum tensor corresponding
to an anisotropic ideal fluid, which we call dark energy.
The two parameters are related to
the coupling of mass to the dark energy and its fall-off as a function of $r$.
The consequences and predictions of this theory will be discussed 
in the context of the observational results of the
{\it Even Horizon Telescope}, expected soon. 
Our main result is that 
due to the accumulation of dark energy near a large mass,
the modified theory predicts 
a dark ring followed by a bright ring
in the emission profile of the accretion disc.
We also discuss the light ring in the equatorial plane.
\end{abstract}




\section{Introduction}
\label{intro}

In our first contribution \citep[][hereafter Paper I]{Boller2019}  we presented observational predictions of a modified theory of Gravity. 
In the present contribution, the theory is explained in greater detail. 

General Relativity (GR) has passed many observational tests \citet{Will2006}
in weak gravitational fields. The first observation of gravitational waves \citet{Abbott2016}
can be considered as a first test in the strong gravitational field. 
Recently bright spots near the black hole in $SrgA^*$ 
were observed \citet{GRAVITY2018} and simulations, assuming that GR is still valid, indicate that they appear close to the {\it Innermost Stable Circular Orbit} (ISCO), 
at six times the gravitational radius.
The {\it Event Horizon Telescope}
\citet{EHT2016} collaboration are expected to publish their results on the observation of SgrA$^*$ 
and the black hole at the center of M87 in 2019.
In \citet{Giddings2018} the oscillations of a black hole are investigated as a consequence of quantum fluctuations. Distortions of the black hole shadow and the shape of a light ring are discussed. They predict that a bright light ring should be present near the event horizon.
In 
\citet{Schoenenbach2013, Schoenenbach2014}, based on \citet{Hess2009}, it 
is assumed that any mass provokes vacuum fluctuations nearby, as 
explained in the book by  \citet{Birrel1982}. 
In \citet{Schoenenbach2013, Schoenenbach2014} the vacuum fluctuations are treated in a 
phenomenological way, assuming a particular increase with 
the mass, such that it is finite at the Schwarzschild radius. 
The vacuum fluctuations are modeled by two parameters, one which describes the coupling of mass with the dark-energy and
the second one on how the dark-energy density falls-off as a function in $r$. Up to now, we assumed the simplest
fall-off behavior such that it does not yet contradict solar system experiment. Recently in \citet{Nielsen2018} the fall-off behavior was restricted with the result that a steeper slope has to be used in order not to contradict gravitational wave observations.

The consequences for the light emission of accretions discs were computed in 
\citet{Schoenenbach2014}. 
The main result is that a dark ring exists in addition to a bright inner one in the emission profile of the accretion disc. It is a consequence of the dependence of the angular velocity of point particles in a circular orbit on the distance and will be explained in the next section. 
This feature is a consequence of vacuum fluctuations, but not of oscillations of a black hole, as proposed in \citet{Giddings2018}. 
The theory in  \citet{Schoenenbach2013, Schoenenbach2014} is static.
Also the recent observation of bright spots \citep{GRAVITY2018} can be reinterpreted such that the circular orbit is near $\frac{4}{3}$ of the gravitational radius ($m$),
for the lowest fall-off and a bit further away for stronger fall-offs.
The theory is phenomenological, it treats the quantum fluctuations as an asymmetric perfect fluid, 
and 
up to now involved only
a parameter describing the coupling of matter to the vacuum fluctuations, also called {\it dark-energy}.
In this contribution we will discuss also stronger fall-offs of the dark energy and show how the structure obtained
will change.
Though it is phenomenological, it may shed some light on the effects of quantization on the standard theory.  
At the end, we will discuss the existence of a light ring in the equatorial plane as a function in $n$, a parameter related to the 
fall-off of the dark energy and which
will be introduced in the next section.

Finally, we point out a different path to extend the standard theory, using canonical transformations \citet{Struckmeier2017}. The Einstein equation has a contribution which represents the known form but also has additional terms which might represent the contribution of the energy-momentum tensor in pc-GR. 

The paper is organized as follows: 
In Section \ref{pcGR} the main ingredients of the extended theory are presented, 
As the point of partition, we present results for the fall-off of the dark energy as done in past publications and 
show how these structures change for varying the $r$-dependence of the dark energy. 
The implications for particles in a circular orbit are discussed as well as the accretion disc emission profiles.
After that, a simulation of a gas clouds falling into the black hole is presented. 
For any emitting material a dark ring followed by a bright one will be one of the main observable results, which will be presented in section \ref{SgrA} for SgrA$^*$.
The properties of the emission profile of the accretion disc as predicted by pc-GR
may be visible in EHT observations.
In this case, the pc-GR theory can be tested directly via the observational tests discussed in Paper I.

\section{The theory, consequences and predictions}
\label{pcGR}

Historically, there have been several attempts to extend GR. 
To our knowledge, \citet{einstein1,einstein2}
was the first to 
extend algebraically the real coordinates in the 4-dimensional space to {\it complex} coordinates, in order to unify Electromagnetism with GR. 
\citet{born1,born2} had different reasons: 
He could not accept that in GR the coordinates play
a dominant role, while in Quantum Mechanics coordinates and momenta are treated on an equal footing. This lead to the
proposal to add to the length element squared 
a term which depends on the momentum of a particle. 
More recently, the 
complex
expanded GR was discussed by \citet{Mantz1,Mantz2}.
However, as shown in \citet{Kelly1986}, only algebraic extensions to the pseudo-complex coordinates (referred to in \citet{Kelly1986} as hyper-complex) make sense, because all others have tachyon or ghost solutions. 
This was the reason why we developed pc-GR.  

In \citet{Hess2009} the pc-GR was introduced, 
which is briefly described in this following section, 
for more details please see for example \citet{Hess2015}: 
The theory extends the real space-time components to the so-called pseudo-complex ones, 
namely $x^\mu$ $\rightarrow$ $X^\mu = x^\mu + I y^\mu$, with $I^2=1$. 
Any function (or set of coordinates) can be expressed in terms of the so-called zero-divisor basis 
$\sigma_\pm = \frac{1}{2}\left( 1 \pm I\right)$, i.e., $F(X)=F_+(X_+)\sigma_+ + F_-(X_-)\sigma_-$ and 
$X^\mu = X_+\sigma_+ + X_-\sigma_-$. 
Because $\sigma_-\sigma_+=0$, the two zero-divisor components commute. 
One obtains the Einstein equations for each component, i.e. $G_{\mu \nu}^\pm - \frac{1}{2}g_{\mu \nu}^\pm R^\pm$
= $ 8\pi T_{\mu\nu}^\pm$ ($c=G=1$). 
In each component the theory appears 
the standard structure of a theory of General Relativity.
All principles and symmetries remain satisfied.  
The connection is obtained requiring that the orbit of a particle is real ($d\omega^2$ is real), which introduces a constraint
whose solution \citep{Hess2017} requires an energy-momentum tensor of an 
an-isotropic ideal fluid. 
In the original version of the pc-GR \citep{Hess2009}, a modified variation principle was proposed, which is not necessary when the constraint is taken into account. 
The theory contains effects of a minimal length scale (maximal acceleration). 
This can be observed when one writes the pc-length element squared, 
using the same functional form of the metric in each zero-divisor component
($g_{\mu\nu}(X_+)$ in the $\sigma_+$-component and $g_{\mu\nu}(X_-)$ in the $\sigma_-$-component), i.e.,
\beqa
d\omega^2 & = & g_{\mu\nu}\left[ dx^\mu dx^\nu + dy^\mu dy^\nu\right] + 2Ig_{\mu\nu}dx^\mu dy^\nu
~~~,
\label{dw2}
\eeqa
where a symmetric metric was assumed. 
This length element has a strong resemblance to the one proposed in \citet{caianello}, when we set $y^\nu = l du^\nu$ ($c=1$), 
with the 4-velocity $u^\nu$. 
When this expression for $y^\nu$ is substituted into the pseudo-imaginary component of (\ref{dw2}) and $dx^\mu$ is
substituted by $\frac{dx^\mu}{d\tau}=u^\mu$ ($\tau$ as the Eigen-time), 
then the pseudo imaginary part is just
$2 l g_{\mu\nu}u^\mu du^\nu$. 
Demanding that this is zero results in the dispersion relation.
However, the identification of $y^\nu$ to the 4-velocity is strictly speaking only valid in a flat space. This is the reason why in pc-GR the coordinate $y^\nu$ is unaltered.

The minimal length parameter $l$ is not noticeable in present day observations and this is why its contributions, encoded in
$y^\mu$, were
neglected in subsequent treatments, which is equivalent to project the Einstein equations to their real part. 
The final equations
have the same structure as the standard equations, save that on the right hand side there must appear an energy-momentum
tensor, which describes an an-isotropic ideal fluid, as mentioned above. The pc-GR cannot determine how this fluid 
has to behave near a central mass and we are dependent on a phenomenological treatment.

The appearance of a non-zero energy-momentum tensor 
is a consequence of the pc-GR formulations and
implies the conjecture
that {\it the mass not only curves space-time but also changes vacuum properties}.
Therefore, in pc-GR the effects of the vacuum fluctuations are associated with the 
the energy-momentum tensor $T_{\mu\nu}$.
In fact, it is shown in semi-classical calculations \citep{Birrel1982,Visser1996}.
that the vacuum fluctuations, with a static metric,
behave proportional to
$1/\left[ r^6\left(1-\frac{2m}{r}\right)^2\right]$, which has an infinity
at the Schwarzschild radius. 
The vacuum fluctuations increase with the strength of the gravitational field and problems arise when it is large, because
vacuum fluctuations are represented by an energy density, which in turn changes the metric. This has to be taken into account.
Because the calculations do not include 
such
back-reaction effects of the vacuum fluctuations to the metric, the behavior near the horizon is not physical and requires
a modification. 
Due to the lack of a functional quantized theory of gravity we are
left to treat 
dark-energy
in a phenomenological way, assuming for 
it
a non-isotropic perfect
and classical fluid 
\citep{Schoenenbach2014}. 

The theory contains two phenomenological parameters, $B_n$ and $n$: The coupling of the dark-energy 
to the mass ($B_n=bm^n$) 
(c.f. Eq. 3)
and its fall-off toward a greater distance, i.e., $\sim 1/r^n$. 
This is the simplest Ansatz and one easily can add further complicated
dependencies in $r$.
Up to now,
the dark energy density was modeled 
by a
$\sim \frac{B}{r^5}$ dependence, 
which falls off strong enough in order not to be detected yet by
solar system observations.

In  \citet{Nielsen2018}, the pc-GR was tested on the results of 
the first observed gravitational wave event \citet{Abbott2016}. 
The corresponding data were not accessible before and are important to restrict the range of the parameters in pc-GR.
Looking at the end of the amplitude evolution of the inspiral event, the pc-GR could not describe it well. Therefore, the
used Ansatz for the $r$-dependence has to be modified. A direct new Ansatz is to increase the power of the fall-off, but
even other dependencies may be used. 
Hence, pc-GR is not excluded!
Therefore, we do study a number of different r-dependencies, as done in \citet{Nielsen2018}. 

The evolution of certain structures, as a function of a parameter $n$, will be explained
starting from a particular $n=3$, 
corresponding to the former ansatz. The main point is that the {\it structures} will be identical
for all $n$, though numerical numbers will change. 
This is what we 
describe as {\it robust}. 
An important objective
is to show, what will happen when dark-energy is accumulated around a large mass and indicate the {\it signatures}.

The parameter $B_n$ is chosen such that the metric component
$g_{00}$ does not become zero, thus no event horizon appears in this theory.  The reasoning for this choice 
is twofold: i) At the Schwarzschild radius the gravitational field is extremely strong, compared to the field in the solar system,
and one has to count on the possibility that GR has to be extended in order to include further
effects (like quantum effects). ii) The second reason is of philosophical nature and depends
on the eye of the beholder, namely that the event horizon in GR 
excludes part 
of the space to be accessible 
to a {\it nearby observer} in the exterior and has several, for us, undesirable 
consequences like the information paradox. 
Both observations indicate 
that maybe GR reaches its limits or rather that something is missing. 
Therefore, our motivation to investigate what happens when there is no event-horizon
for a so-called black hole.
 
Recent presentations of pc-GR are given in 
\citet{Schoenenbach2014,Hess2017}
who apply pc-GR to thin, optically thick accretion discs. 
Though a thin optical disc is not always realized (except perhaps around SgrA$^*$), 
the effects discussed in this contribution apply equally to other disc models, 
which is also important for the black hole in the center of M87.
The modified Kerr metric is given by 
\citet{Schoenenbach2014}, \citet{Schoenenbach2014c}.


\beqa
g^{{\rm K}}_{00} &= \frac{ r^2 - 2m r  + a^2 \cos^2 \vartheta + \frac{B_n}{(n-1)(n-2)r^{n-2}}} 
{r^2+ a^2 \cos^2\vartheta} \nonumber \\
g^{{\rm K}}_{11} &= - \frac{r^2 + a^2 \cos^2 \vartheta}{r^2 - 2m r + a^2 +  \frac{B_n}{(n-1)(n-1)r^{n-2}}  } \nonumber \\
g^{{\rm K}}_{22} &= - r^2 - a^2 \cos^2 \vartheta  \nonumber \\
g^{{\rm K}}_{33} &= - (r^2 +a^2 )\sin^2 \vartheta - \frac{a^2 \sin^4\vartheta \left(2m r -  \frac{B_n}{(n-1)(n-2)r^{n-2}}  \right)}{r^2 + a^2 \cos^2 \vartheta}  \nonumber \\
g^{{\rm K}}_{03} &= \frac{-a \sin^2 \vartheta ~ 2m r + a \frac{B_n}{(n-1)(n-2)r^{n-2}}   
\sin^2 \vartheta }{r^2 + a^2 \cos^2\vartheta}   \quad ,
\label{eq:kerrpseudo}
\eeqa
where the index K refers to {\it Kerr} (in \citet{Schoenenbach2012,Hess2015} the $B_n$ is divided by an extra factor 2, which should
not be there). For $B=0$ the Kerr solution of GR is obtained.
The $n$ is the above mentioned parameter, which is related to the one  in \citet{Nielsen2018}
$n_{\rm N} = n-1$.
There is no event horizon for
\beqa
B_n & > & \frac{2(n-1)(n-2)}{n}\left[ \frac{2(n-1)}{n}\right]^{n-1}m^n ~=~ B_{\rm max}
~~~,
\label{B}
\eeqa
which for $n=3$ is $\frac{64}{27}m^3$ and for $n=4$, the next possible value, it is $\frac{81}{8}m^4$.
For the equal sign in (\ref{B}) there would be a horizon
at the position
\beqa
 r_0=\frac{2(n-1)}{n} m
~~~,
\label{r0}
\eeqa
which for 
$n=3$ acquires the value $r_0=\frac{4}{3}$~m and for $n=4$ it is $r_0=\frac{3}{2}$~ m.  
The rotational parameter $a$ is in units of $m$ and ranges as usual from $0$ to $1$~m. 
Often the rotational parameter in observations is redefined without units.

\begin{figure}
\centerline{
\rotatebox{180}{\resizebox{280pt}{200pt}{\includegraphics[width=0.60\textwidth,angle=90]{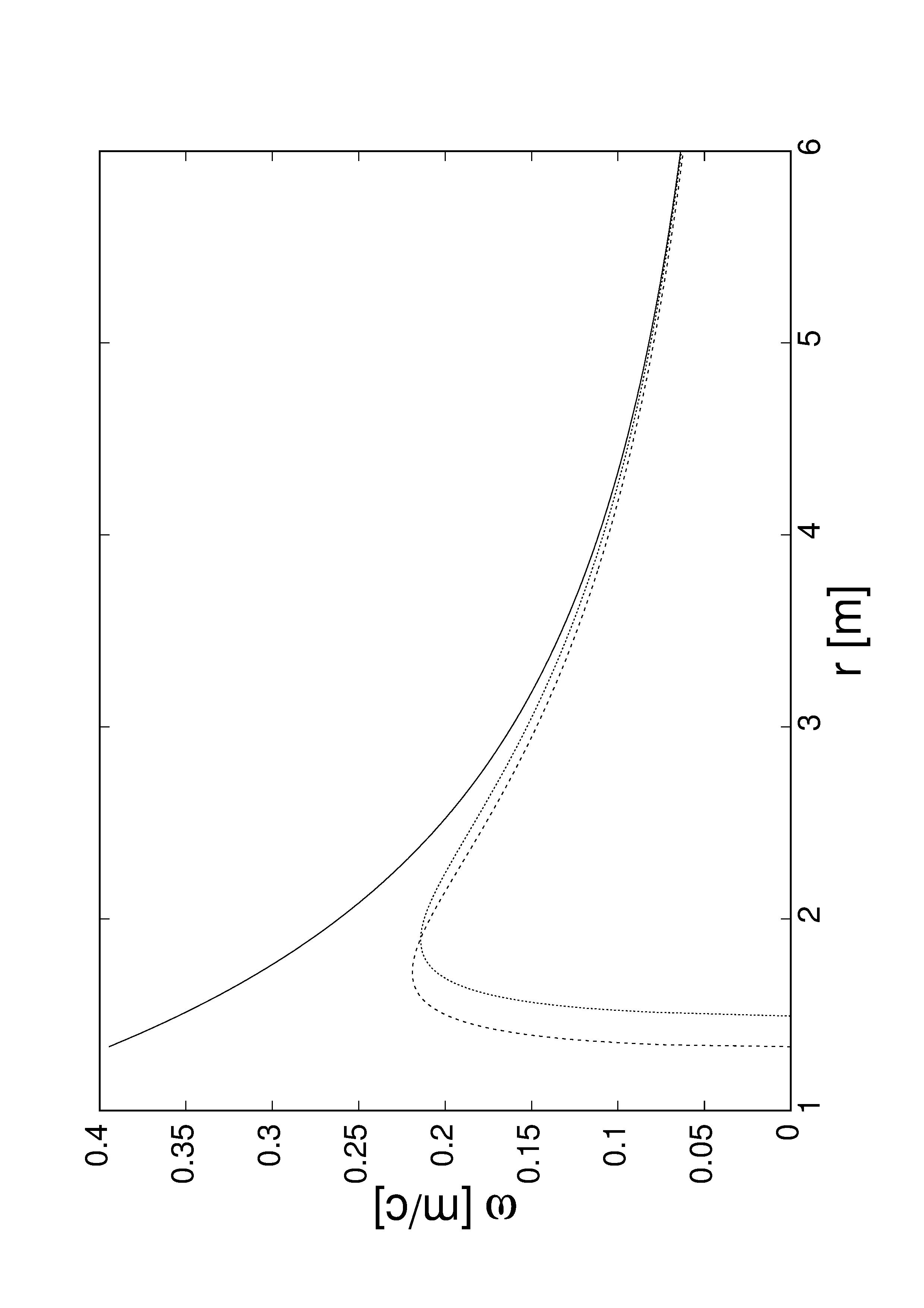}}}}
\caption{ The orbital frequency of a particle in  a circular orbit for the case GR (upper curve) and for $n=3$ (dashed curve) and $n=4$ (dotted curve) $B=B_{\rm max}$.
\label{om34}
}
 \label{orbits}
\end{figure}

The orbital frequency for a particle in a circular orbit and for $n=3$ was given in \citet{Schoenenbach2013,Hess2015}.
For $n$ arbitrary and for prograde orbitals, this is
\beqa
\omega_n &=& \frac{1}{a+\sqrt{\frac{2r}{h_n(r)}}} ~,~ h_n(r) = \frac{2}{r^2} - \frac{nB_n}{(n-1)(n-2)r^{n+1}}
~,
\eeqa
For $B_n$ given by the right-hand expression in (\ref{B}), the $h_n(r)$ acquire the form
\beqa
h_3(r) & = & \frac{2}{r^4}\left(r^2 - \frac{16}{9}\right) ~,~ h_4(r) ~ = ~ \frac{2}{r^5}\left(r^3 - \frac{27}{8}\right)
~~~,
\eeqa
In Fig. \ref{om34} the orbital frequency for $n=3$ and $n=4$,  for $a$=0.995, is compared to GR.

One distinct feature of the modified theory is that the orbital frequency
is always lower than in GR, though noticeable differences only appear near the Schwarzschild 
radius 
(\citet{Schoenenbach2014}.
Furthermore, the orbital frequency shows for $B_{\rm max}$ 
a maximum at 
\beqa
r_{\omega_{\rm max}} & = & \left[ \frac{n(n+2)b_{\rm max}}{6(n-1)(n-2)}\right]^{\frac{1}{n-1}}~{\rm m}
~~~,
\label{rom}
\eeqa
which is independent of the value of $a$,
after which it falls off toward the
center and reaches zero at $r_0$ (Eq. (\ref{r0})).
For $n=3$ the value is approximately $r$=1.72~ m and for $n=4$ it is $r$=1.89~m.

Taking these numbers one notes that the zero of the orbital frequency {\it and} the position of its
maximum move further out with increasing $n$.
The fall off 
of the orbital frequency towards smaller 
$r$
is explained by the increase of the dark energy, which effectively reduces the 
gravitational constant (it is a function in $r$, see also \citet{Hess2015}). In
\citet{Schoenenbach2014} it is shown that for $\rm a\lesssim0.4~m$ 
($n=3$)
the last stable orbit is at a lower value of $r$ but follows approximately the behavior 
of GR.
Because particles can now reach further inside, more gravitational energy is released, and as a consequence the disc appears to be brighter. 
Above $a=0.4~m$ stable orbits are possible up to the surface.
Circular stable orbits are possible up to the maximum 
of the orbital frequency. Because at the maximum the
gradient in the orbital frequency of neighboring orbits is small, less friction is
present, resulting in less heating. Thus, a dark ring appears near the position 
$r_{\omega_{\rm max}}$, which is 1.72 for $n=3$. 
Further inside, the gradient becomes significantly stronger and a bright ring occurs 
(see Paper I). 

What will change as a function in $n$ is the position of the dark  and brighter inner ring. For small $a$
the pc-GR curve approaches the ISCO of GR and the point from which on there are stable orbits in pc-GR up
to the surface approaches larger values. The main point is that the {\it signatures} of the dark and bright
ring will not change.
Note, that these features are independent on the detailed structures of the disc model.
If a dark ring is observed, 
its position can be used to determine the optimal value of $n$
using Eq.~\ref{rom}.
%
We use the model proposed by \citet{Page1974} for the accretion disc.
It assumes a thin disc which is optically thick at the equator
and the loss of energy is only through the emission of light.
However, we acknowledge that other accretion disc models may also be valid. 
The model exploits the laws of conservation of mass, energy, and angular momentum. 
Properties such as the viscosity must be implicitly included
in the parameters of the model (for example the accretion rate). 
This has to be extracted from somewhere else. 

Though the model was applied exclusively to GR up to now,
it is easy to extend it to other metrics, such as the one used in pc-GR. 
This is because the procedure discussed in \citet{Page1974} 
is formulated in such a way that it is independent of the kind of metric that is used.
For the simulation we use the GYOTO routine \citet{Vincent2011}, which applies
the ray-tracing method and allows the use of several models, 
including that used by Page and Thorne, permitting
an arbitrary metric. 
In \citet{Vincent2011}
the theoretical background of the ray tracing technique is clearly described.

The basic concepts are that light-rays
are incident on the detector. 
which is divided into pixels. 
The light rays are then {\it followed back}
to the accretion disc, 
such that only rays coming from there are taken into account.
In order to construct the paths of the photons a Hamilton-Jacobi method is applied and
the so-called {\it Carter constant} is constructed, which is a conserved quantity, which has to 
be checked at various points along the path in order to minimize numerical errors. 
These equations also involve the metric and, thus, the details of the
simulations will depend on the metric. 
The path of a photon depends on the metric and the
 intensity profile also depends on the radius of the last stable orbit, which also depends indirectly on the metric.
The smaller the radius of the last stable orbit, the more energy can be released. 

It is also possible to change the subroutines in GYOTO to be consistent with the model of
\citet{Page1974}.
Such an extension for pc-GR is available in
\citet{Schoenenbach2014b}.
When the model of \citet{Page1974} is not applicable
and other models have to be used, the
differences to GR, 
nevertheless, will be similar in structure because the effects only depend on the orbital frequency as a function in $r$.

Under the  hypothesis that,
for $n=3$ the,
$B=\frac{64}{27}m^3$, simulations of accretion discs have been 
performed in 
\citet{Schoenenbach2014,Hess2015}. However, in these works, 
the total flux  was calculated in {\it bolometric units}, 
the mass was normalized to the mass of the black hole in the Galactic Center,
and the accretion rate was set to an arbitrary value.
The EHT on the other hand observes well defined objects 
at a fixed wavelength of approximately 1.2~mm which corresponds to a frequency of 250~GHz 
\citet{EHT2016}. 
Also the field of view was chosen arbitrarily 
and the temperature of the disc near the inner edge has not been evaluated.

The results of the simulation depends on several well determined parameters, such as the 
accretion rate, the rotational parameter $a$, the field of view, and the mass of the
black hole. 
The specific uncertainties for these parameters are calculated in relative units.
Nevertheless, the important part is the {\it relative comparison} to GR. 
In pc-GR the intensities are much larger than in GR (see Paper I).
We have performed simulations from $10^o$ to $80^o$ in steps of ten degrees.
One such simulation is presented in Paper I, Figure 1.
%
For the remainder of this section, we discuss the position of the light ring, in the equatorial plane) as a function of $n$. 
The position of the light ring is obtained demanding that the path of a photon (infinitesimal length element squared
put to zero) is also geodesic \citet{Schoenenbach2012,Nielsen2018}. This corresponds to demand that two 
frequencies, one for the circular orbital geodesic orbit and for the path of the photon in a circular orbit, have to be set equal.
We used the Eqs. (7) and (14) in \citet{Schoenenbach2013}, with
now $n$-dependent metric coefficients, and Eqs. (7) and (9) in \citet{Nielsen2018} and compared the 
results for consistency. For $B_n$ we used the value given in (\ref{B}).
The properties we found are as follows: 
i) From a minimal value, $a_{\rm min}>0.8~m$ on until $a=1~m$,
the condition for the existence of the light ring provides no real positive solution for $r$, i.e, no
light ring is possible
ii) For $n=3$ there is a light ring up to approximately 
$a=0.95$~m, starting at approximately $r=$ 2.93~m at $a=0$
and tending to lower $r$ values for increasing $a$. For larger $a$ the point up to which a light ring exists is lowered systematically, reaching for $n=8$ the value approximately 0.77~m. Whether this is relevant for SgrA$^*$ cannot be concluded
because the lower limit for $a$ in SgrA$^*$   at 0.5~m \citet{Genzel2003}.  
\section{SgrA$^*$}
\label{SgrA}
%
%
%
%
%
%
\begin{figure}
\centerline{
\rotatebox{270}{\resizebox{200pt}{270pt}{\includegraphics[width=0.23\textwidth]{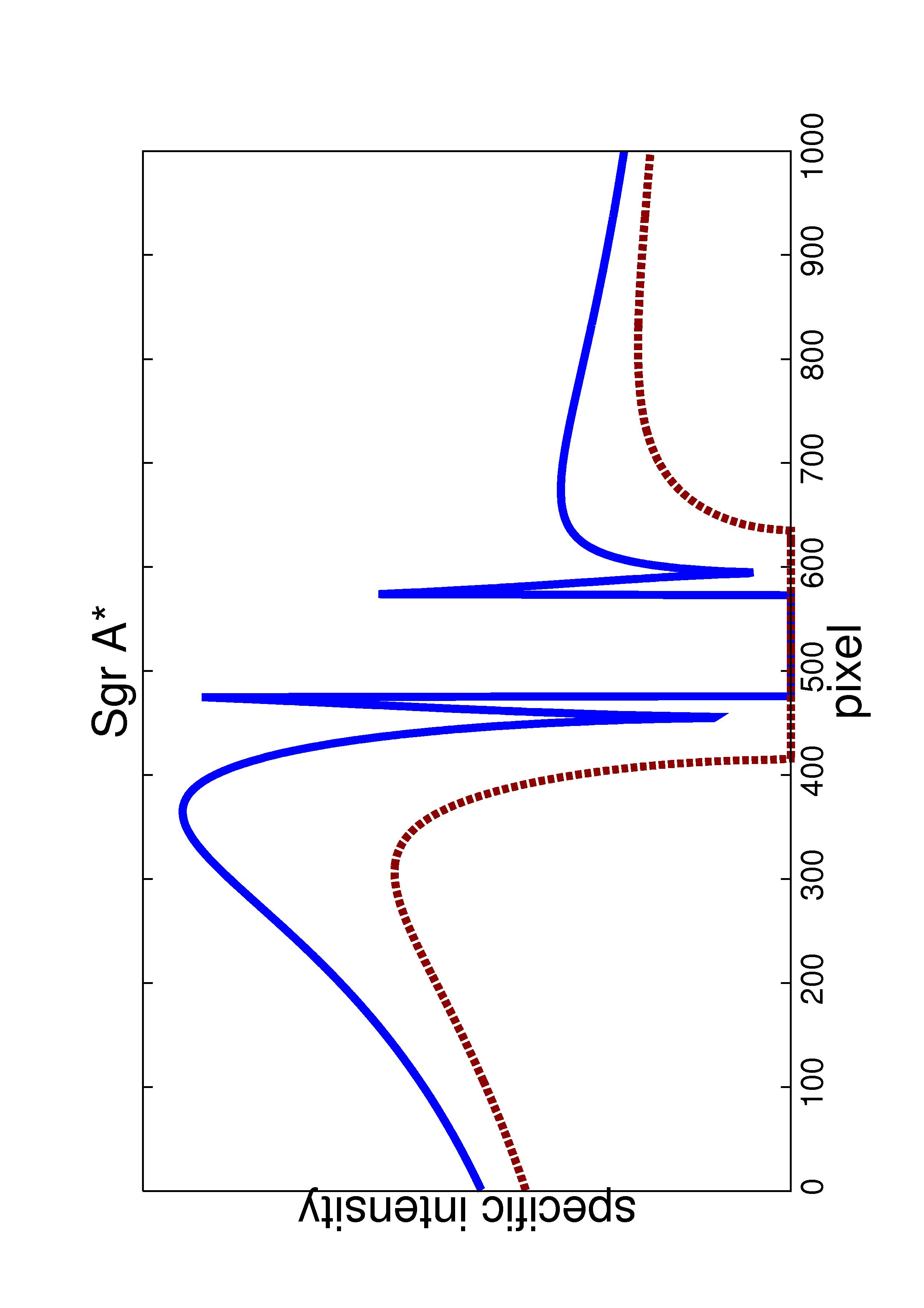}}} 
}
\caption{ 
\color{black}
Comparison between standard GR (dashed red line) 
and pc-GR (solid blue line) of the intensities
for SgrA$^*$
($n=3$, $a=0.9$~m). 
Scales are not shown.
The inclination angle was set to 80$^o$. 
A dark ring is seen at zero intensity, while
the bright ring is represented by a large peak further in. 
Due to resolution effects in the plot, the inner curve does not 
acquire an exact zero.
For $a=0.9m$, standard GR does not show sharp peaks in the predicted intensity. 
For pc-GR, the inner bright ring goes sharply to zero, approaching the surface. 
The horizontal axis corresponds to the range of approximately 22 m.
}
 \label{SgrA80curva}
\end{figure}
The main parameters for Sgr A$^*$ are its mass and the accretion rate. 
The mass of the black hole is known to be
about 4.3 million solar masses
\citep{Gillesson2009}
and an upper limit of the accretion rate is estimated in
\citet{Quataert1999} with the value of ${\rm 8}\times 10^{-5}$ solar masses per year. Under the 
supposition that an accretion disc forms around SgrA$^*$, the same value for the
accretion rate will be used. For SgrA$^*$ there is no proof for the existence of an accretion disc. 
In the case where no accretion disc exists, 
when a cloud approaches the black hole it will pass through regions where an
accretion disc would show different intensities. 
We expect that the changes in intensity will
be qualitatively reflected when the cloud approaches the black hole. 
We assume a spatial resolution of 
0.1~$\mu {\rm as}$
over a field of 1000 pixels, resulting into
a field of view of 100~$\mu {\rm as}$.
The simulations provide figures for the {\it specific intensity} at the fixed frequency mentioned above. The unit for the specific intensity is ${\rm erg}~{\rm cm}^{-2}{\rm Hz}^{-1}{\rm sr}^{-1}$
(GYOTO manual 2015).
The version of  the ray tracing routine we used, 
provided by \citet{Vincent2011},
developed numerical instabilities for very large distances. Because the specific intensity is independent of the distance, the simulations were performed at a distance of
1000$m$. 
The field of view was correspondingly adjusted, which for 
SgrA$~*$ at a distance of 1000 $m$ is about 0.022~rad.
Intensive ray-tracing simulations for the black hole in the Galactic Center have been obtained by 
\citet{Schoenenbach2014} for different spin and inclination values. 
%
Note, that the intensity in pc-GR is significantly larger than
in GR. Also, at a distance (about $1.72~m$
for $n=3$, for larger values this position changes according to the discussion in the text) 
a dark ring occurs, which is a consequence of the
maximum of the orbital frequency of a particle in a circular orbit 
\citet{Schoenenbach2013}.
Near this distance, friction is low (and therefore viscosity) and the emission of light reaches a 
minimum. 
This is also noted in Figure \ref{SgrA80curva}, where the intensity
is plotted versus the pixel number. The distance between two pixels corresponds to approximately 0.022m.
The intensity reaches a minimum at a distance
of $1.72~m$ and for lower radial distances the intensity increases again, which corresponds to the
bright ring shown in 
Fig. 1 in paper I \citet{Boller2019}. 
In GR this behavior is not seen. 

\section{Conclusions}
We summarized the basics of a modified theory of General Relativity, 
which depends on two phenomenological
parameters, namely the coupling of the dark energy to the mass ($B_n$) and 
the power of
the fall off of the dark energy as 
a function of the radial distance.
One of the main predictions is that the accretion disc emission will 
show both an outer dark ring and an inner bright ring. 
Though a thin accretion disc was assumed in this work,
the results we obtained are robust and can also be applied for other types of emission. 
This is because these features depends only on the $r$-dependence of the orbital frequency 
of particles in a circular orbit.
Another main difference is that in pc-GR the
specific intensity is larger than in GR (see also the related discussion in Paper I). 
The reason for this is that 
stable orbits exist at smaller radii,
and therefore, more energy is released.
A further prediction is that the surface 
is at approximately
$r_0=\frac{2(n-1)}{n}~m$, where $n$ is a parameter of
the theory which describes the rise of the dark-energy density toward the center. This parameter can be
restricted by gravitational wave analysis \citet{Nielsen2018} and by observations of the {\it Event Horizon Telescope} \citet{EHT2016}.
 i.e. in the Schwarzschild case ($a=0$)
the size of the black hole is smaller than in the standard theory
and in the Kerr case with $a=1m$ it is larger. 
In addition, for the spin parameter $a$, below $a=1$ there is a range of values where
there is no light ring possible. This range  is for $n=3$ between 0.95 and 1 and increases its size for larger $n$.
%

These predictions and others (see also Paper I) should be verifiable by the {\it Event Horizon Telescope} and they are the same for the black hole in the center of M87.
\section*{Acknowledgements}
TB and POH are grateful for the longstanding support and collaboration with the Frankfurt Institute of Advanced Studies. 
The authors thank Damien Coffey for a critical reading of the manuscript.
POH acknowledges financial support from 
{\it Programa de Apoyo a Proyectos de Investigaci\'on e Innovaci\'on Tecnol\'gica} (PAPIIT) (IN100418).
H.St. acknowledges the Judah M. Eisenberg Professor Laureatus at the Fachbereich Physik endowed by the Walter Greiner-Gesellschaft zur F\"orderung der physikalischen Grundlagenforschung e.V. Frankfurt am Main.
The authors would like to thank the referee whose comments greatly improved the content of this paper.







\bsp	
\label{lastpage}

\begin{thebibliography}{99}
\bibitem[\protect\citeauthoryear{Boller}{2019}]{Boller2019}
Boller T., Hess P. O., M\"uller, A., St\"ocker, H., 2019 MNRAS 485, 34


\bibitem[\protect\citeauthoryear{Abbott}{2016}]{Abbott2016}
Abbott B. P.  et al. (LIGO Scientific Collaboration and Virgo Collaboration), 2016, Phys. Rev. Lett., 116 , 061102


\bibitem[\protect\citeauthoryear{Birrell}{1982}]{Birrel1982}
Birrell, N. D., Davies P. C. W., 1982, {\it Quantum Fields in Curved Space} (Cambride University Press, Cambridge)

\bibitem[\protect\citeauthoryear{Bonning}{2007}]{Bonning2007}
Bonning E. W., Cheng L., Shields G. A. S., Salvander S., Gebhard K., 2007, ApJ 659, 211

\bibitem[\protect\citeauthoryear{Born}{1948}]{born1} Born M., 1938,  Proc. Roy. Soc. A 16, 291.

\bibitem[\protect\citeauthoryear{Born}{1949}]{born2} Born M., 1949, Rev. Mod. Phys. 21, 463.

\bibitem[\protect\citeauthoryear{Caianiello}{1981}]{caianello} Caianiello E. R., 1981, Nuovo Cim. Lett. 32, 65.

\bibitem[\protect\citeauthoryear{EHT}{2016}]{EHT2016} EHT, 2016, http://www.eventhorizontelescope.org

\bibitem[\protect\citeauthoryear{Einstein}{1945}]{einstein1} Einstein A., 1945, The Ann. of Math. 46,
578.

\bibitem[\protect\citeauthoryear{Einstein}{1948}]{einstein2} Einstein A., 1948, Rev. Mod. Phys. 20,
35.

\bibitem[\protect\citeauthoryear{Genzel}{2003}]{Genzel2003}
Genzel R., Sch\"odel R., Ott T., et al., 2003, Nature 425, 934G


\bibitem[\protect\citeauthoryear{Giddings}{2018}]{Giddings2018}
Giddings S. B., Psaltis D., 2018, arXiv:1606.07814[astro-ph]


\bibitem[\protect\citeauthoryear{Gillessen}{2009}]{Gillesson2009}
Gillessen, S., Eisenhauer, F., Trippe,  Alexander, T., Genzel, R., Martins, F., Ott, T.,
2009, ApJ 692, 1075


\bibitem[\protect\citeauthoryear{GRAVITY}{2018}]{GRAVITY2018}
GRAVITY Collaboration, R. Abuter et al., 2018,  A\&A 618, 201834294

\bibitem[\protect\citeauthoryear{GYOTO}{2015}]{GYOTO2015}
GYOTO manual, 2015, http://www.gyoto.obspm.fr/GyotoManual.pdf

\bibitem[\protect\citeauthoryear{Hess}{2009}]{Hess2009}
Hess P. O., Greiner W., 2009, Int. J. Mod. Phys. E18, 51

\bibitem[\protect\citeauthoryear{Hess}{2015}]{Hess2015}
 Hess P. O., Sch\"afer M., Greiner W., {\it  Pseudo-Complex General Relativity},
(Springer, Heidelberg, 2015).

\bibitem[\protect\citeauthoryear{Hess}{2016}]{Hess2016}
Hess P. O., 2016, MNRAS 462, 3026.

\bibitem[\protect\citeauthoryear{Hess}{2017}]{Hess2017}
Hess P. O. and W. Greiner, Centennial of General Relativity: A Celebration, Edited
by: C. A. Zen Vasconcellos (World Scientifiic Publishing, Singapore, 2017), p. 97.

\bibitem[\protect\citeauthoryear{Kelly}{1986}]{Kelly1986}
Kelly P. F., Mann R. B., 1986, Class. Quantum Grav., 3, 705

\bibitem[\protect\citeauthoryear{Kluzniak}{2007}]{Kluzniak2007}
Kluzniak W. and Rappaport S., 2007, ApJ, 671, 1990

\bibitem[\protect\citeauthoryear{Mantz}{2008}]{Mantz1}  Mantz C. L. M., Prokopec T., 2008, 
{\it  Hermitian Gravity and Cosmology}, arXiv:0804.0213

\bibitem[\protect\citeauthoryear{Mantz}{2011}]{Mantz2} Mantz C. L. M., Prokopec T., 2011,
Phys. 41, 1597.

\bibitem[\protect\citeauthoryear{Nielsen}{2018}]{Nielsen2018}
Nielsen A., Birnholz O., 2018, AN 339, 298

\bibitem[\protect\citeauthoryear{Page}{1974}]{Page1974}
Page D. N., Thorne K. S., 1974, ApJ, 191, 499

\bibitem[\protect\citeauthoryear{Quataert}{1999}]{Quataert1999}
Quataert E., Narayan R.,  Reid M. J., 1999, The Astroph. J. 517, L101 

\bibitem[\protect\citeauthoryear{Sch{\"o}nenbach}{2012}]{Schoenenbach2012}
Caspar G., Sch\"onenbach T., Hess P. O., Sch\"afer M., Greiuer W., 2012
Int. J. Mod. Phys. E 21, 1250015. 

\bibitem[\protect\citeauthoryear{Sch{\"o}nenbach}{2013}]{Schoenenbach2013}
Sch\"onenbach T., Caspar G., Hess P.O., Boller T., M\"uller A., Sch\"afer M., 
Greiner W., 2013, MNRAS 430, 2999

\bibitem[\protect\citeauthoryear{Sch{\"o}nenbach}{2014}]{Schoenenbach2014}
Sch\"onenbach T., Caspar G., Hess P.O., Boller T., M\"uller A., Sch\"afer M., 
Greiner W., 2014,  MNRAS 442, 121

\bibitem[\protect\citeauthoryear{Sch{\"o}nenbach}{2014b}]{Schoenenbach2014b}
Sch\"onenbach T, 2014b, https://github.com/schoenenbach/Gyoto

\bibitem[\protect\citeauthoryear{Sch{\"o}nenbach}{2014c}]{Schoenenbach2014c}
Sch\"onenbach T, 2014c, PhD thesis, Universit\"at Frankfurt am Main, Germany

\bibitem[\protect\citeauthoryear{Schoedel}{2002}]{Schoedel2002}
Sch\"odel R., Ott T., Genzel R. et al., 2002, Nature 419, 694

\bibitem[\protect\citeauthoryear{Struckmeier}{2017}]{Struckmeier2017}
Struckmeier J., Muench J., Vasak D., Kirsch J., Hanauske M. and Stoecker H., 2017, Phys. Rev. D 95, 124048

\bibitem[\protect\citeauthoryear{Vincent}{2011}]{Vincent2011}
Vincent F. H. Paumard T., Gourgoulhon E., Perrin G., 2011, Class. 
Quantum Grav. 28, 225011

\bibitem[\protect\citeauthoryear{Visser}{1996}]{Visser1996}
Visser M., 1996, Phys. Rev. D 54, 5116

\bibitem[\protect\citeauthoryear{Will}{2006}]{Will2006}
Will C. M., Living Rev. Relativ., 2006, 9, 3

\end{thebibliography}
\end{document}